\documentclass[prl,aps,reprint,superscriptaddress]{revtex4-1}
\usepackage{bm,amsmath,epsfig,color}
\newcommand{\rev}[1]{\textcolor{black}{{#1}}}

\begin{document}

\title{Glassy behaviour of sticky spheres: What lies beyond experimental timescales?}

\author{Christopher J. Fullerton}

\affiliation{Laboratoire Charles Coulomb (L2C), Universit\'e de Montpellier, CNRS, 34095 Montpellier, France}

\affiliation{Department of Physiology, Anatomy and Genetics, University of Oxford, Oxford, United Kingdom}

\author{Ludovic Berthier}

\affiliation{Laboratoire Charles Coulomb (L2C), Universit\'e de Montpellier, CNRS, 34095 Montpellier, France}

\affiliation{Department of Chemistry, University of Cambridge, Lensfield Road, Cambridge CB2 1EW, United Kingdom}

\date{\today}

\begin{abstract}
We use the swap Monte Carlo algorithm to analyse the glassy behaviour of sticky spheres in equilibrium conditions at densities where conventional simulations and experiments fail to reach equilibrium, beyond predicted phase transitions and dynamic singularities.
We demonstrate the existence of a unique ergodic region comprising all the distinct phases previously reported, except for a phase-separated region at strong adhesion.
All structural and dynamic observables evolve gradually within this ergodic region, the physics evolving smoothly from well-known hard sphere glassy behaviour at small adhesions and large densities, to a more complex glassy regime characterised by unusually-broad distributions of relaxation timescales and lengthscales at large adhesions.
\end{abstract}

\maketitle

Steeply repulsive particles with very short-range attractive forces (`sticky spheres') are experimentally realised with colloids~\cite{hunter2012physics,gonzalez2016colloidal}.
When the attraction range is small compared to the particle size, the physics of sticky spheres differs qualitatively from that of atomic liquids~\cite{baxter1968percus,noro2000extended,sciortino2002one}.
Sticky spheres thus represent a unique paradigm for the statistical mechanics of soft materials and simple fluids, motivating a large number of theoretical studies and experiments.
The phase diagram of sticky spheres is explored by changing the volume fraction and the adhesion strength, showing interesting behaviour at low (clustering and phase separation~\cite{post1986cluster}) and large (crystallisation~\cite{bolhuis1997isostructural,lee2008effect}, glassy dynamics~\cite{pham2004glasses}) volume fractions.

Over the last two decades, the glass transition of sticky spheres received considerable attention.
This effort gathered momentum when the mode-coupling theory (MCT) of the glass transition~\cite{gotze2008complex} was applied to the square-well potential to predict the phase behaviour and glassy dynamics of sticky spheres~\cite{bergenholtz1999nonergodicity,dawson2000higher,dawson_2001,gotze2003higher,sperl_2004}.
The predicted existence of two types of glass transition, of reentrant glassy dynamics, and of a glass-glass phase transition line ending at a singular critical point giving rise to non-trivial relaxation patterns triggered massive theoretical~\cite{geissler2005short,sellitto2013thermodynamic,sellitto2013packing,ghosh2019microscopic,ghosh2020microscopic},  numerical~\cite{zaccarelli2001mechanical,puertas_2002,zaccarelli_2002,foffi2002evidence,zaccarelli_2003,sciortino_2003,zaccarelli_2004,saika_voivod_2004,reichman2005comparison,moreno_2006} and experimental~\cite{mallamace2000kinetic,pham2002multiple,eckert2002re,pham2004glasses,kaufman2006direct,buzzaccaro_2007,lu_2008,zhang2011cooperative,whitaker2019colloidal} efforts, which continue to this day.  

\rev{For this reason,} published work is often torn between successes and failures of these MCT predictions. Two recent computational studies~\cite{zaccarelli_2009,royall_2018} offer contradicting conclusions even on basic features of glassy sticky spheres and important physical questions are left unanswered \rev{which go well beyond the relevance of MCT predictions.} \rev{There is a broad agreement on the existence of reentrant dynamics along isochores~\cite{pham2002multiple,foffi2002evidence,pham2004glasses}, non-trivial dynamic correlation functions at intermediate adhesion~\cite{zaccarelli_2002,sciortino_2003}, and increasingly localised particle motion at large adhesion~\cite{pham2004glasses}.} On the other hand the \rev{existence, nature and physical relevance} of the MCT liquid-glass and glass-glass lines, of various phases (equilibrium gel, attractive, repulsive, bonded and non-bonded glasses), and the interplay between gelation, glassiness and phase separation remain debated. Resolving these questions \rev{has been technically too difficult so far,} as large relaxation timescales plague both computer simulations and experiments, and prohibit the exploration of the equilibrium phase diagram. Informative non-equilibrium aging studies at large densities have been performed instead~\cite{foffi2004aging,zaccarelli_2004,zaccarelli_2009}. 

Here we show that the swap Monte Carlo algorithm, which has recently provided an equilibration speedup larger than $10^{11}$ in several three-dimensional model glass-formers~\cite{ninarello_2017} (including hard spheres~\cite{berthier_2016,coslovich2018local,berthier2019bypassing}), performs equally well for dense sticky spheres.
This decisive computational advance allows us to perform a complete exploration of the equilibrium phase diagram of sticky spheres, including regions at large densities where distinct phases were predicted or numerically reported. Our simulations instead reveal the existence of a broad ergodic fluid phase limited at large adhesions by a phase-separated region where non-equilibrium gelation may occur. Within the ergodic fluid, the dynamics is reentrant along isochores, and evolves smoothly between the well-known hard sphere limit to a more complex sticky glassy dynamics characterised by a broad hierarchy of relaxation timescales and lengthscales, but this appears distinct from the predicted MCT phases and singularities, which we do not observe.   

We describe sticky spheres using the well-studied system of hard spheres decorated with a short-range attractive square-well.
Particles separated by $r_{ij}$ have interaction energy $V(r_{ij} \leq \sigma_{ij}) = \infty$, $V(\sigma_{ij} < r_{ij} < \lambda\sigma_{ij}) = -u$, and $V( r_{ij} > \sigma_{ij}) = 0$ where $(\lambda-1)\sigma_{ij}$ defines the width of the attractive well, and $\sigma_{ij} = (\sigma_i+\sigma_j)/2$.
To compare with other studies, we use the relative width $\epsilon=(\lambda-1)/\lambda = 0.03$ of the square well~\cite{zaccarelli_2002,zaccarelli_2003,zaccarelli_2004}.
This value is often used as for it MCT predicts the existence of an $A_3$ singularity within the glass phase, close enough to affect the system's dynamics at points where it can be equilibrated on accessable timescales.
We use a continuous distribution of particle diameters, $P(\sigma_{\mathrm{min}} \leq \sigma \leq \sigma_{\mathrm{max}}) = A/\sigma^3$, where $A$ is a normalisation constant.
We choose $\sigma_{\mathrm{min}}$ and $\sigma_{\mathrm{max}}$ to provide a polydispersity of $\Delta = \sqrt{\langle\sigma^2\rangle - \langle \sigma \rangle^2}/ \langle \sigma \rangle = 23\%$.
This choice \rev{prevents crystallisation, makes the swap Monte Carlo algorithm efficient, and does not appear to lead to novel features in the dynamics, when compared to different types of size dispersity~\cite{berthier_2016,ninarello_2017,PhysRevLett.125.085505}.} We use Monte Carlo dynamics to explore the structure and dynamics of the system. Equilibration is achieved using swap Monte Carlo, with details as in \cite{ninarello_2017,fullerton_2017}.
To analyse the dynamics, we perform conventional Monte Carlo simulations, which describe glassy dynamics equivalently to Brownian and Molecular Dynamics~\cite{berthier2007monte}.
We simulate $N=1000$ particles in a periodic cubic box of volume $V$.
The packing fraction is $\phi = \pi N \langle \sigma^3 \rangle/ (6V)$.
We fix the temperature $\beta = k_B T = 1$ and vary the well depth and packing fraction to explore the $(u,\phi)$ phase diagram.
\rev{This produces equivalent results to varying the temperature 
with $u$ fixed, as $u/T$ is the appropriate control parameter.}
Additional simulations with $N=8000$ are performed to investigate the phase separation boundary at large $u$.
Times are measured in units of Monte Carlo steps, where a step represents $N$ attempted Monte Carlo moves (swap or translational), and distances in units of the average particle diameter $\langle \sigma \rangle$.

To quantify dynamics, we calculate the mean-squared displacement (MSD) defined as $\langle r^2(t) \rangle = (1/N) \sum_i |{\bf r}_i(t) - {\bf r}_i(0)|^2$, where ${\bf r}_i(t)$ is the position of particle $i$ at time $t$.
The MSD is the second moment of the van Hove distribution of single particle displacements: $G_s(x,t) = \langle \delta (x -|x_i(t)-x_i(0) | ) \rangle$, for displacements along the $x$-direction (later averaged over all directions).
We define the self-part of the incoherent scattering function: $f({q}, t) = (1/N) \sum_j e^{i {\bf q}. ( {\bf r}_j(t) - {\bf r}_j(0))}$.
We perform a spherical average at $|{\bf q}| = 7.8$, close to the first peak of the static structure factor, and define the structural relaxation time $\tau_{\alpha}$ as $f(|{\bf q}|=7.8, \tau_{\alpha}) = e^{-1}$.
When the system is nearly arrested, we fit these functions using $f(q,t) = f_q + h_q [B_q^{(1)}\ln(t/\tau) +B_q^{(2)}\ln^2(t/\tau)]$~\cite{puertas_2002, sciortino_2003}, mainly to extract the non-ergodicity parameter $f_q$. 

To ensure efficient equilibration at large $\phi$, we use swap Monte Carlo. At each state point, we define $\tau^{\rm swap}_\alpha$ via $f(q,t)$ measured in the presence of swap moves.
Note that all particles (small and large) need to relax for this function to decay, which ensures full ergodicity.
We consider our system as adequately equilibrated if it has been simulated longer than $4 \tau_{\alpha}^{\rm swap}$~\cite{ninarello_2017}.
We collect independent equilibrium configurations at many state points $(u,\phi)$ to study static behaviour, and from these we launch many independent, conventional Monte Carlo simulations lasting up to $t_s = 5 \times 10^8$ MC steps to analyse the equilibrium dynamics over a broad time window, including at conditions where the physical relaxation time $\tau_\alpha$ is \rev{larger than $t_s$ by many orders of magnitude.} This is only possible thanks to the combined use of swap and conventional Monte Carlo. 

\begin{figure}
\includegraphics[width=8.5cm]{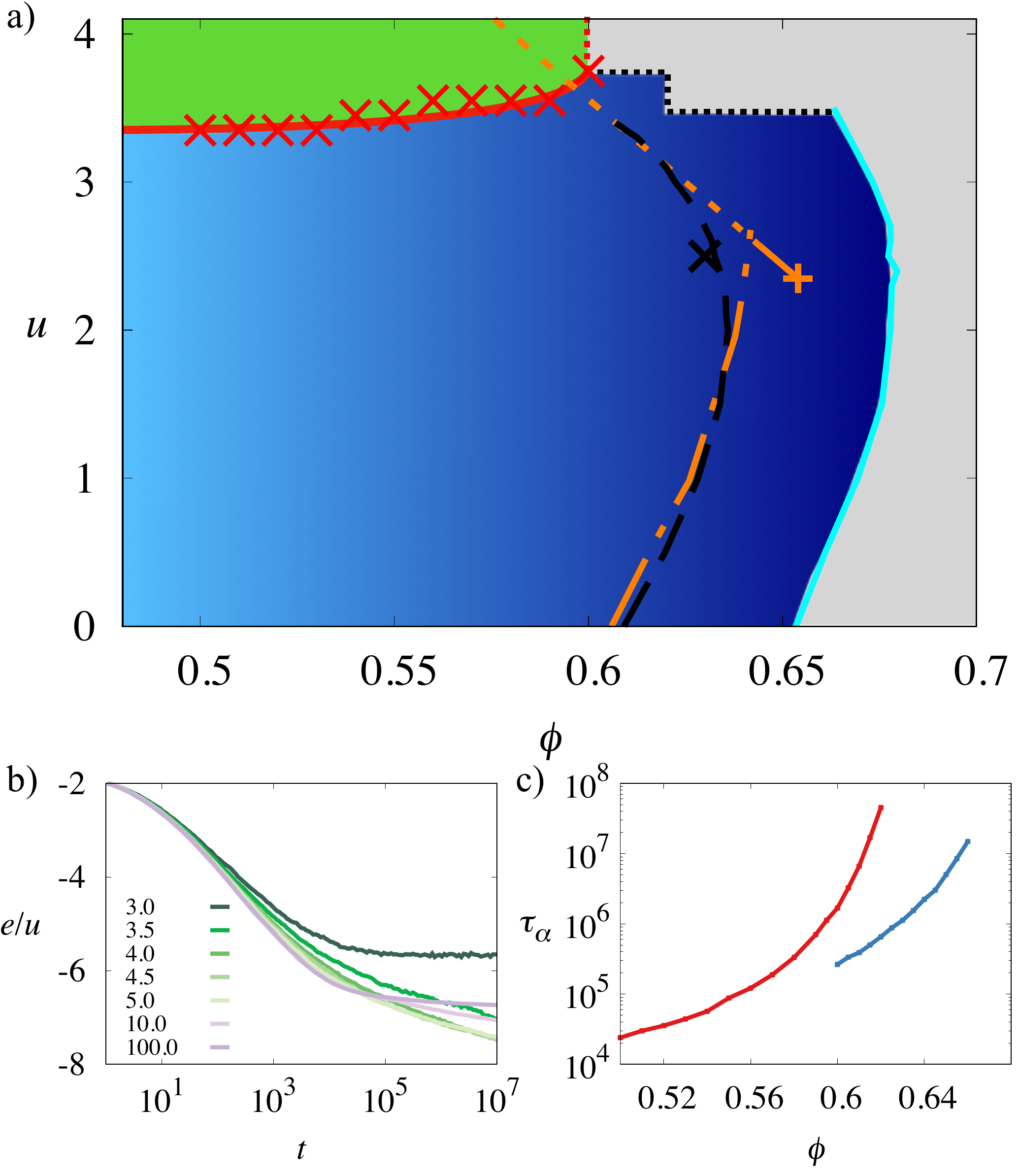}
\caption{a) Equilibrium phase diagram $(u,\phi)$ of sticky spheres, with a large ergodic fluid (blue) and a phase separated region (green). The isochrone 
	$\tau^{\mathrm{swap}}_{\alpha} = 10^7$~MC steps (full cyan line) limits the ergodic region at large $\phi$, whereas the isochrone $\tau_{\alpha} = 10^7$~MC steps (black dashed line) marks the limit of conventional simulations. The avoided MCT singularities are mapped in orange \rev{(dashed: fluid to attractive glass, dot-dashed: fluid to repulsive glass, solid: glass-glass line)} ending at the $A_3$ singularity \rev{(orange symbol)}. Grey areas could not be explored in equilibrium, and the black cross corresponds to Fig.~\ref{fig:logdecay}(a). b) Potential energy as a function of the time after a quench along the isochore $\phi=0.5$. c) Relaxation times with (blue) and without (red) swap at $u = 3.0$.}
\label{fig:PD}
\end{figure}

The decisive progress provided by the swap algorithm can be appreciated in Fig.~\ref{fig:PD}(a), which shows the equilibrium $(u,\phi)$ phase diagram.
We distinguish two regions.
The large blue area comprises state points where we achieved thermal equilibrium.
This region extends to arbitrarily low $\phi$, and is limited at large $u \gtrsim 3.5$ by a phase-separated region. The ergodic region is limited at large $\phi$ by our ability to reach equilibrium, \rev{i.e. by the time spent running simulations. With more time, or an algorithm more efficient than swap we expect the ergodic region to extend to higher $\phi$.}
We empirically define the right-most boundary as the isochrone where $\tau_\alpha^{\rm swap} =10^7$~MC steps, but densities even larger than this $\phi \approx 0.65-0.66$ empirical boundary could presumably be explored by performing longer simulations. 

There is a single phase transition in Fig.~\ref{fig:PD}(a), \rev{not described by MCT.} Holding $\phi$ constant and increasing $u$, the system phase separates into two phases with distinct densities.
The potential energy after quenching from $u=0$ to different points along the isochore $\phi=0.5$ is shown in Fig.~\ref{fig:PD}(b).
The large jump and different time dependence of the energy at long times between $u = 3$ and \rev{$u = 3.5$ indicates phase separation occurs somewhere between these points. The heterogeneous structure of the system was studied at long times, and the red crosses in Fig.~\ref{fig:PD}(a) placed at points above which phase separation is seen to have occured. This can be done up to large volume fractions near $\phi=0.6$~\cite{royall_2018} but at larger $\phi$, the amount of low-density phase becomes too small and the coarsening too slow to identify the phase separation clearly. At very large $u$, the system resembles a gel but the slow decrease in energy at long times shows that the system coarsens~\cite{testard2011influence}, indicating the gel is not stable.}

Everywhere in the blue area of Fig.~\ref{fig:PD}(a), the system is an ergodic fluid. The significance of this conclusion comes when considering the physical dynamics of the system.
Increasing $\phi$ at constant $u$, the relaxation time $\tau_{\alpha}$ increases very fast and the system becomes arrested on the observational timescales, as shown for $u=3.0$ in Fig.~\ref{fig:PD}(c).
This figure also illustrates the giant speedup afforded by swap Monte Carlo at high $\phi$ for sticky particles.
We report in Fig.~\ref{fig:PD}(a) the isochrone $\tau_\alpha =10^7$~MC steps, which marks the limit where conventional simulations equilibrate.
The isochrones with and without swap are parallel, but separated by a large gap $\Delta \phi \simeq 0.05$, nearly independent of $u$.
\rev{This wide new territory is explored in equilibrium for the first time here, and provides distinct insight into the physics of sticky spheres at large densities.}

Crucially, all distinct phases reported previously for this system belong to the same ergodic fluid phase.
We conclude that none of these phases actually exists as such, and the phase diagram is much simpler than anticipated~\cite{dawson2000higher,pham2002multiple,zaccarelli_2009} with only two phases separated by the well-known discontinuous liquid-gas thermodynamic instability.
Deep inside the phase separating region, coarsening towards a fully demixed state may become slow, but is not arrested~\cite{testard2011influence}.
Given the time window accessible to colloidal experiments, the phase separation is never complete and the system behaves as a colloidal gel~\cite{foffi2005arrested,manley2005glasslike,lu_2008,zaccarelli2008gelation}, with physical properties that are slowly aging.
This represents the non-equilibrium route to colloidal gelation~\cite{zaccarelli2007colloidal}.     

The ergodic region contains \rev{in particular} all sharp features theoretically predicted by MCT which we map in Fig.~\ref{fig:PD}(a) by following earlier work fitting our measured relaxation times to MCT power law predictions. The liquid-glass and glass-glass transition lines ending at the $A_3$ singularity all belong to the ergodic fluid.
Therefore, they represent, at best, smooth physical crossovers~\cite{ghosh2019microscopic}. \rev{Our demonstration that all ideal MCT singularities disappear in physical systems of sticky spheres echoes equivalent findings for molecular glasses~\cite{berthier2011theoretical} and colloidal hard spheres~\cite{brambilla2009probing}, which remain debated in the colloidal context~\cite{van2010comment,reinhardt2010comment,brambilla2010reply,PhysRevLett.105.199605,zaccarelli2015polydispersity}. }

\begin{figure}
\includegraphics[width=8.5cm]{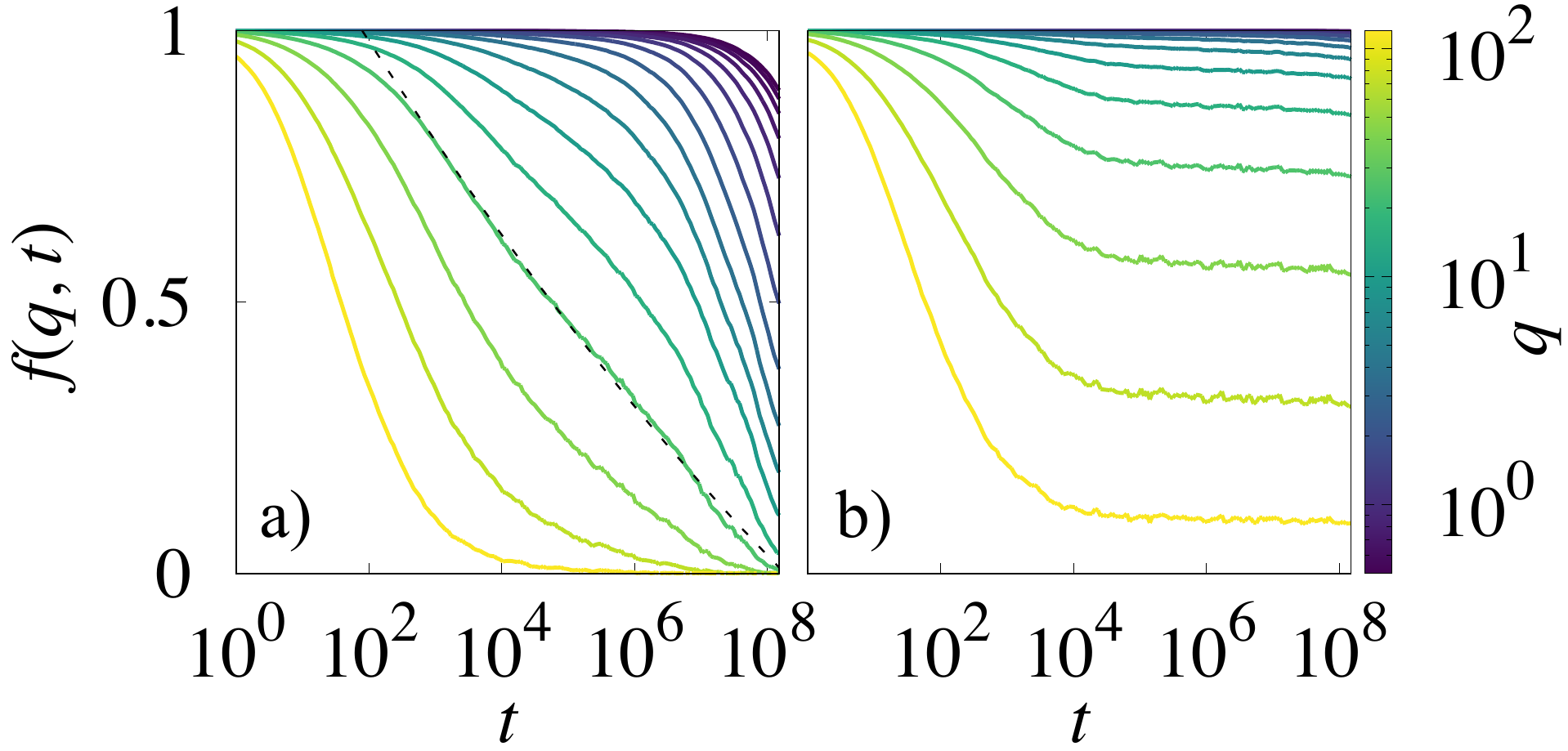}
\caption{$f(q,t)$ at various $q$ measured at a) the black cross $(\phi=0.630,u=2.5)$ or b) the orange cross $(\phi = 0.654, u=2.5)$  in Fig.~\ref{fig:PD}a. The near-logarithmic decay highlighted in a) for $q=26.3$ is no longer present closer to the putative $A_3$ singularity in b).}
\label{fig:logdecay}
\end{figure}

Is the concept of an \rev{{\it avoided}} $A_3$ singularity \rev{nevertheless} useful? Our model displays the physical behaviour expected for a system with competing attractive and repulsive interactions.
The banana-shaped iso-$\tau_\alpha$ line in Fig.~\ref{fig:PD}(a) implies reentrant glassy dynamics as $u$ varies along isochores.
Reentrance is mathematically described by MCT via the existence of two distinct glass transition lines, but these are not required to explain it~\cite{ghosh2019microscopic}.
Much less trivial is the observation of a transient `logarithmic' decay of $f(q,t)$ at well-chosen state points approaching the $A_3$ point~\cite{zaccarelli_2002,sciortino_2003} where the relaxation should become purely logarithmic~\cite{gotze2003higher}.
In Fig.~\ref{fig:logdecay}(a), we show $f(q,t)$ at $(\phi =0.630,u=2.5)$ (black cross in Fig.~\ref{fig:PD}(a)) for a range of wavevectors $q$.
The decay time increases with decreasing $q$, showing that the system remains mobile on short length scales but is frozen on long length scales.
At intermediate $q$-values, a nearly logarithmic time dependence holds over about 5 decades, a behaviour clearly distinct from the conventional two-step decay observed in most glassy materials~\cite{berthier2011theoretical}. 

Previous work attributed this unsual dynamics to proximity to the $A_3$ singularity~\cite{foffi2002evidence,zaccarelli_2002,sciortino_2003}.
We can test this hypothesis directly by measuring the equilibrium dynamics much closer to the $A_3$ singularity, as in Fig.~\ref{fig:logdecay}(b).
We find that all hints of logarithmic behaviour are gone, the dynamics now being consistent with a simpler two-step decay.
(At timescales much larger than those shown here, structural relaxation will eventually take place.)
These data suggest that the existence of an $A_3$ singularity may not be the best physical way to interpret the unconventional dynamics in Fig.~\ref{fig:logdecay}(a).
It was shown, for instance, that by numerically tuning the strength of competing attractive and repulsive interactions~\cite{zaccarelli_2003,chaudhuri2010gel,chaudhuri2015relaxation}, a near-logarithmic decay may appear or disappear, or be replaced by a simpler multi-step decay. 

Our results dispel the possibility that several distinct phases characterize dense sticky spheres~\cite{pham2002multiple,zaccarelli_2009}.
No sharp distinction exists between attractive, repulsive, bonded and non-bonded glasses. Instead, we now show that increasing adhesion smoothly changes the physics between two qualitatively-distinct types of glassy dynamics.
To see this, we explore the large $\phi$ region using several paths in the phase diagram changing either $u$ or $\phi$. 

\begin{figure}
\includegraphics[width=8.5cm]{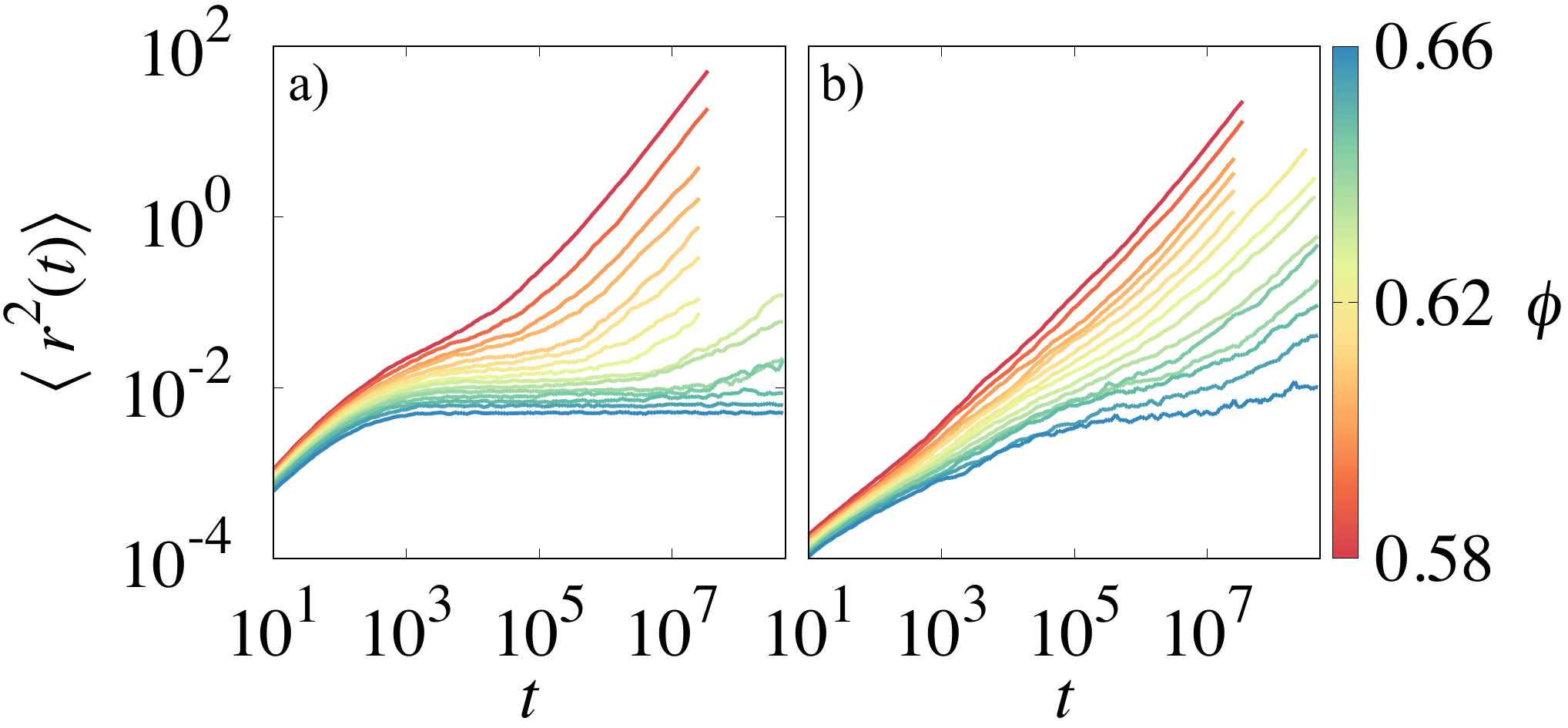}
\caption{Evolution of the MSD with packing fraction for a) $u=0.$ and b) $u  =3.0$. The equilibrium glassy physics at short timescales and lengthscales for the adhesive system is different from that of hard spheres.}
\label{fig:MSD}
\end{figure}

Glassy dynamics is encountered for any $u \lesssim 3.5$ as $\phi$ is increased, see Fig.~\ref{fig:MSD}.
In all cases, the diffusion constant drops by several orders of magnitude as $\phi$ increases, until diffusion becomes too slow to be observed.
However, interesting differences can be seen between repulsive and sticky particles.
When $u=0$, the MSD displays a well-defined plateau, whose amplitude decreases smoothly with $\phi$.
For $u=3.0$ no well-defined plateau can be seen, even for packing fractions as large as $\phi=0.66$ (remember that all data are taken in equilibrium \rev{and so we do not expect a plateau at even larger times}).
The plateau is replaced by a slow subdiffusive regime that extends over 7 decades in time \rev{dramatically distinct from the hard sphere cage physics.}
This behaviour also differs from ideas of an MCT-inspired attractive~\cite{dawson2000higher,zaccarelli_2003} or a bonded~\cite{zaccarelli_2009} glass \rev{suggested from simulations,} and is not to be confused with non-equilibrium gelation either~\cite{royall_2018}.

\begin{figure}
\includegraphics[width=8.5cm]{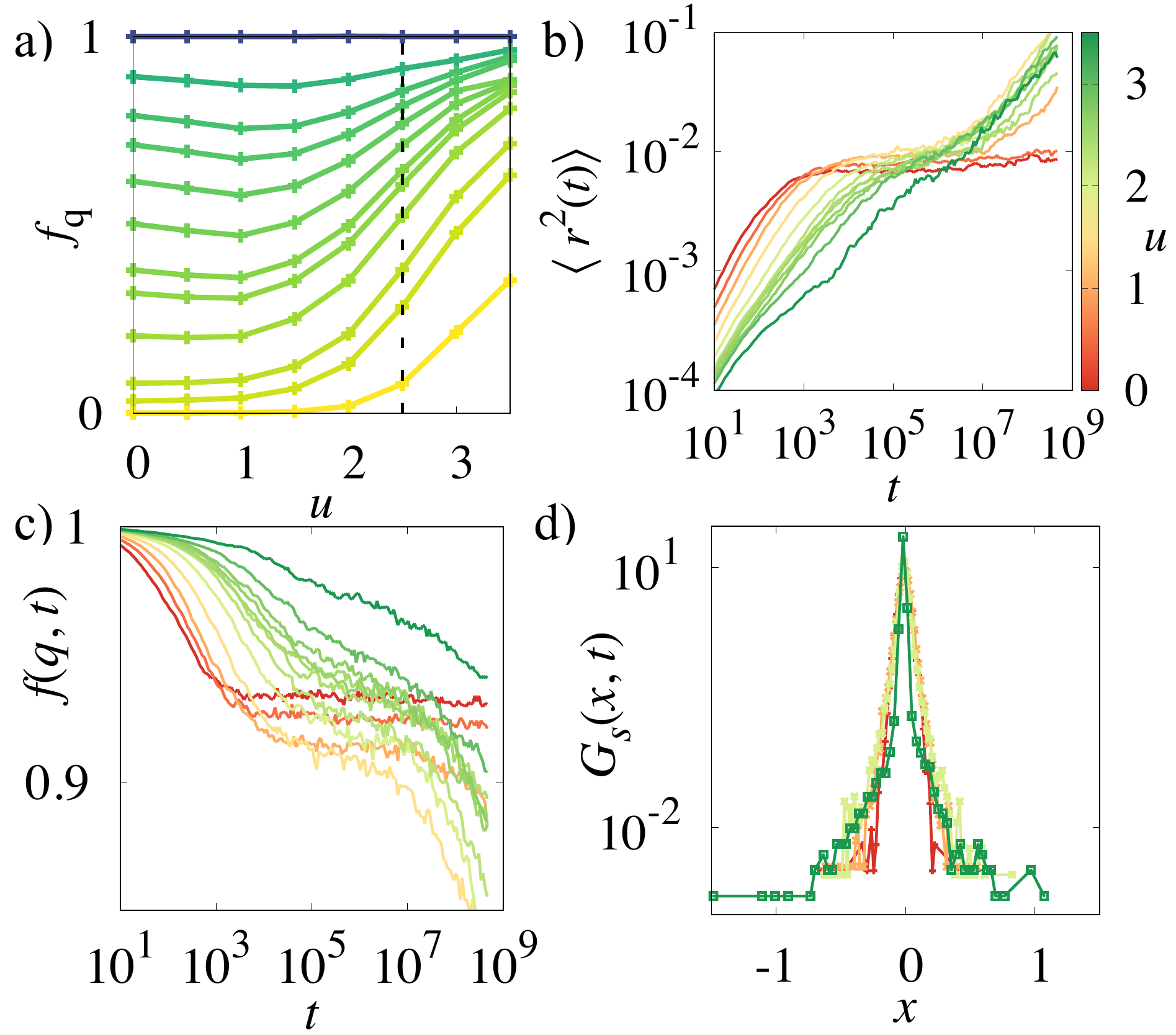}
\caption{Evolution of a) the non-ergodicity parameter, b) the mean-squared displacement, c) the intermediate scattering function, d) the van Hove function along the isochore $\phi=0.65$ at equilibrium.} 
\label{fig:nonerg_and_MSD}
\end{figure}

The sharp distinction between attractive and repulsive glasses is nonexistent, but in the regime $u \approx 2.5-3.5$ between phase separation and hard spheres the system exhibits unusual glassy dynamics, \rev{uncovered here thanks to swap Monte Carlo.} We characterize this regime further in Fig.~\ref{fig:nonerg_and_MSD} by changing $u$ along the $\phi=0.65$ isochore, which crosses the (putative) glass-glass line very close to the $A_3$ singularity.
This isochore lies in the region where equilibration can only be achieved using swap.
 
In Fig.~\ref{fig:nonerg_and_MSD}(a), we show the non-ergodicity parameter.
At all wavevectors $f_q$ is higher at $u = 3.5$ than it is at $u = 0$, showing that stronger adhesion means less mobility at all lengthscales.
The change in $f_q$ is greatest for large $q$ (short lengthscales).
When $u$ is small, particles are free to move within the hard sphere cages but are immobilised on long lengthscales.
As $u$ increases, the attractive well can trap (or `bond'~\cite{zaccarelli_2009}) particles at much shorter distances.
Attractive interactions also destabilise the hard sphere glass, which results in a slight non-monotonic behaviour of $f_q$ at small $q$ near $u=1.5$.
Again, $f_q$ varies smoothly with $u$ (this is true across a range of $\phi$) in contrast to the sharp jump predicted across the MCT glass-glass line. 

The marked (but gradual) evolution along the $\phi=0.65$ isochore is further illustrated in Figs.~\ref{fig:nonerg_and_MSD}(b,c) showing the time dependence of $\langle r^2(t) \rangle$ and $f(q,t)$.
These functions change dramatically in the range $u \in [0, 3.5]$.
At small $u$ a well-developed plateau exists: the particles are caged by repulsive interactions with their neighbours.
The approach to this long-lived (6 decades in time) plateau is fast.
As $u$ increases clear signs of a structural relaxation speedup appear at long times, together with a weakening of the plateau.
Increasing $u$ further the fast approach to a plateau gets replaced by a slow sub-diffusion (in $\langle r^2(t) \rangle$), or a slow decay (in $f(q,t)$).
This shows that at large $u$ particles are neither caged nor bonded, but instead get arrested over multiple lengthscales, ranging from very short corresponding to the attractive well width to larger than the hard sphere cage size, which is no longer relevant.
This differs from the picture of a bonded glass~\cite{zaccarelli_2009}, but leaves room for a glass transition where adhesion is relevant, at odds with \cite{royall_2018}.
Rather they demonstrate that the structure and short-time dynamics of sticky spheres at large $u$ is highly heterogeneous~\cite{reichman2005comparison,kaufman2006direct,zhang2011cooperative}, and involves a very broad hierarchy of timescales and lengthscales long before structural relaxation. 

The increasing heterogeneity of the glassy structure of sticky spheres is finally confirmed by the evolution of the van Hove distribution in Fig.~\ref{fig:nonerg_and_MSD}(d).
A near-Gaussian distribution is observed at small $u$, confirming the pertinence of a description of the hard sphere glass with a typical cage size~\cite{charbonneau2012dimensional}.
By contrast the van Hove distribution is much broader and strongly non-Gaussian at large $u$, with both a large peak at very small displacements and a fat non-Gaussian tail at large displacements, suggesting enhanced dynamic heterogeneity~\cite{reichman2005comparison}.  

Using swap Monte Carlo, we have explored the complete equilibrium phase diagram of dense sticky spheres.
A \rev{clarifying physical} picture emerges with three distinct regimes of slow dynamics. At large adhesions, $u \geq 3.5$, the system phase separates at least up to $\phi=0.60$ and discontinuously enters a slowly coarsening aging regime leading to non-equilibrium gelation.
At small $u \leq 1.5$ and large $\phi$ the system displays well-known hard sphere glassy dynamics, characterised by a two-step decay of correlation functions and a well-defined cage size at intermediate times.
Finally, in the regime $u=1.5-3.5$ and large $\phi$ unusual glassy dynamics are observed, characterised by a broad distribution of relaxation timescales and length scales and a short-time dynamics \rev{distinct} from hard spheres.
We are aware of no atomic or molecular experimental analog of this unsual glassy behaviour, which involves multiple (time and length) scales and extended sub-diffusion long before the structural relaxation.
The sharp distinction predicted by MCT between two types of glassy dynamics is invalidated by the data, which \rev{more importantly} do not support the physical relevance of an avoided $A_3$ singularity to interpret the dynamics.
The transient logarithmic time decay has a simpler interpretation and is not seen on approaching the $A_3$ location.
The very unsual time correlation functions we report are instead observed at a much larger adhesion strength, away from the avoided $A_3$ singularity.
The proposed clarification of the phase behaviour and dynamics of dense sticky systems should help reinterpreting past experiments and suggest new ones.
Future numerical work could also help understand better the rheological behaviour~\cite{zaccarelli2001mechanical,pham2006yielding,pham2008yielding,altieri2018microscopic,moghimi2020mechanisms} in adhesive colloidal glasses, \rev{by subjecting high density equilibrium states to quasi-static and oscillatory shear.}

\begin{acknowledgments}
We thank M. Cates, P. Royall and E. Zaccarelli for useful exchanges. This work was supported by a grant from the Simons Foundation (Grant No. 454933, L. B.).
\end{acknowledgments}

\bibliography{sticky_spheres}

\end{document}